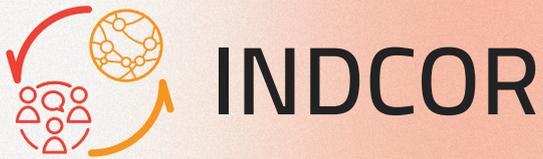

# INDCOR

INDCOR WHITEPAPER 3

# Interactive Digital Narratives and Interaction

**AUTHORS**

Frank Nack[1], Sandy Louchart[2], Kris Lund[3],
Mattia Bellini[4], Iva Georgieva[5],
Pratama W. Atmaja[6], Peter Makai[7]

[1] INDElab, Informatics Institute, University of Amsterdam (nack@uva.nl)
[2] School of Simulation and Visualisation, The Glasgow School of Art (s.louchart@gsa.ac.uk)
[3] CNRS, Interactions, Corpus, Apprentissage, Représentations (ICAR) Research lab, Ecole Normale Supérieure de Lyon (kristine.lund@ens-lyon.fr)
[4] Research Group on Narrative, Culture, and Cognition, University of Tartu (mattia.bellini@ut.ee)
[5] Philosophy of Science Department, Institute of Philosophy and Sociology, Bulgarian Academy of Sciences (ivavgeorgieva@gmail.com)
[6] Department of Informatics, University of Pembangunan Nasional "Veteran" Jawa Timur, Indonesia (pratama_wirya.fik@upnjatim.ac.id)
[7] Kazimierz Wielki University, Bydgoszcz, Poland (peter.makai@mensa.hu)

**AUTHORS**

Frank Nack[1], Sandy Louchart[2], Kris Lund[3],
Mattia Bellini[4], Iva Georgieva[5],
Pratama W. Atmaja[6], Peter Makai[7]

[1] INDElab, Informatics Institute, University of Amsterdam (nack@uva.nl)
[2] School of Simulation and Visualisation, The Glasgow School of Art (s.louchart@gsa.ac.uk)
[3] CNRS, Interactions, Corpus, Apprentissage, Représentations (ICAR) Research lab, Ecole Normale Supérieure de Lyon (kristine.lund@ens-lyon.fr)
[4] Research Group on Narrative, Culture, and Cognition, University of Tartu (mattia.bellini@ut.ee)
[5] Philosophy of Science Department, Institute of Philosophy and Sociology, Bulgarian Academy of Sciences (ivavgeorgieva@gmail.com)
[6] Department of Informatics, University of Pembangunan Nasional "Veteran" Jawa Timur, Indonesia (pratama_wirya.fik@upnjatim.ac.id)
[7] Kazimierz Wielki University, Bydgoszcz, Poland (peter.makai@mensa.hu)

**INDCOR WHITEPAPER 3**

# Interactive Digital Narratives and Interaction

This article/publication is based upon work from COST Action INDCOR, CA18230, supported by COST (European Cooperation in Science and Technology).

COST (European Cooperation in Science and Technology) is a funding agency for research and innovation networks. Our Actions help connect research initiatives across Europe and enable scientists to grow their ideas by sharing them with their peers. This boosts their research, career and innovation.

www.cost.eu

# Table of Contents

# Executive Overview

The nature of interaction within Interactive Digital Narrative (IDN) is inherently complex. This is due, in part, to the wide range of potential interaction modes through which IDNs can be conceptualised, produced and deployed and the complex dynamics this might entail. The purpose of this whitepaper is to provide IDN practitioners with the essential knowledge on the nature of interaction in IDNs and allow them to make informed design decisions that lead to the incorporation of complexity thinking throughout the design pipeline, the implementation of the work, and the ways its audience perceives it.

This white paper is concerned with the complexities of authoring, delivering and processing dynamic interactive contents from the perspectives of both creators and audiences.

This white paper is part of a series of publications by the INDCOR COST Action 18230 (Interactive Narrative Design for Complexity Representations), which all clarify how IDNs representing complexity can be understood and applied (INDCOR WP 0 – 5, 2023).

# Introduction

The aim for this whitepaper on Interactive Digital Narrative (IDN) and interaction is to provide readers with a set of foundational knowledge, definitions and concepts that allow them to develop a deeper understanding of the factors enabling meaningful interactions within a dynamic narrative framework. This whitepaper is focused on providing readers with pointers, information and resources towards relevant avenues of research and practice to anyone with an interest in IDNs or interaction. For a detailed discussion of the advantages and challenges of IDNs in the context of representing complex issues the reader is referred to the INDCOR White Paper 0 (INDCOR WP 0, 2023).

In essence, an IDN is a form of interactive media, leaning towards the idea of a narrative environment in the form of a cybernetic system (Koenitz, 2023), where the observed outcomes of actions are taken as inputs for further action. Interactions drive the design of IDN technologies, tools and production methodologies, as the specificities of user engagement have to be integrated at the core of IDN concepts and development. The audience's understanding of an IDN is a direct product of user interaction with the interactive narrative artefact.

The whitepaper comprises, therefore, three distinct but related sections. Part I focuses on the definitions and concepts of interaction that contribute to the conceptualization of IDN. Part II considers the main processes of IDN authoring (addressing the point of view of the creator), meaning making (the point of view of the audience), and impact (the point of view of the discourse). Part III provides a critical reflection on the introduced concepts in the context of perceptions, communications and the still existing challenges and overall impact of IDNs.

# Part I. Interaction (Definitions, concepts)

Precise definitions of interaction depend on the vantage point of the person doing the looking (Longino, 2013). More generally, this means that each discipline's or community's way of thinking about concepts, methods, and theories is influenced by constraints on how to view the world, on what research questions people find interesting, and on how to go about answering them (Lund et al, 2020).

There are many disciplines and communities that have worked on the notion of interaction, which are relevant for the EU COST Action on "Interactive Narrative Design for Complexity Representations" (INDCOR). The scope of this whitepaper does not permit us to include all such definitions, but we have chosen a small subset, given our own expertise. We examine chosen definitions of interaction from the viewpoint of Interactional Linguistics, Embodied Interaction, Cognitive Science, Human Computer Interaction (HCI), and Interactive Digital Narrative (IDN). The consequences of how defining interaction in a particular way influences the decisions regarding IDN design, implementation, use and further exploration made by stakeholders (researchers, journalists, game designers, etc.) will be made in Part II and III of this paper.

## Interactional linguistics

Interactional linguistics focuses on the use of language within social interaction and although researchers represent diverse traditions, their "... unifying perspective is to describe linguistic structures and meanings as they serve social goals in naturally occurring spoken, in a broad sense, conversational language, viz. 'talk-in-interaction'" (Lindström, 2009, p. 96). Language forms and practices are constantly adjusting to their context and thus contribute to the emergence of aspects that are relevant to the context (Lund, 2019). More recently interactional linguists approach concepts on a level of micro-analytical specifics of how humans co-construct their embodied interactions from the point of view of contextualised language practices (Morek, 2015). For example, Lund et al (2022) argue that language is a complex adaptive system, examining its place in relation to interactive, pragmatic, multimodal discourse processes, but also in relation to cognition, argumentation and meaning-making, and to social structures and education. Considering an IDN as a communication system that contextualises relations between the self and other, private and public, inner thought and outer world, the outlined epistemological assumptions by interactional linguists are in general relevant for the explanation of role of interaction in IDN and can be found in various incarnations of IDNs and related theoretical work (see INDCOR WP 3, 2023).

## Embodied interaction

Streeck, Goodwin, & LeBaon (2011) note that the variety with which the organisation of action in human interaction can be investigated. Whereas other disciplines may look at the mental intentions of individual actors or alternatively, at large, historically shaped social structures, they choose to study "events in which multiple parties are carrying out endogenous courses of action in concert with each within face-to-face human interaction" (opt.cit., p. 1). Put another way — with a more extensive focus — by Mazur & Traverso (2022):

> "In simple terms (if we dare in this context), interactive processes are complex first and foremost because, if considered as composite systems, they involve a very large number of elements: resources related to languages (syntax, lexicon, words, sounds, etc.) as well as to other semiotic fields (gestures, gaze, face expression, manipulation of objects and artefacts); different senses (sight, hearing, touch, smell); contexts, activities and actions; objectives (that can be local, global, and that evolve as the exchanges unfold); stakes of different levels; participants, to whom are attached numerous possible characterisations, such as ongoing social relations, identities, cultures, emotions, etc."

Some aspects of the above definition are taken into consideration by drama related forms of IDN (Mateas & Stern, 2005; Peinado et al, 2008; Aylett et al, 2011). With respect to game-oriented IDNs or newer developments, like narratives in the metaverse, embodied interaction is naturally relevant.

## Cognitive Science

Cognitive science is the interdisciplinary, scientific study of the mind that examines the nature, the tasks, and the functions of cognition (in a broad sense). Cognition here covers the mental action or process of acquiring knowledge and understanding through thought, experience, and the senses. Cognitive scientists study intelligence and behaviour, with a focus on how nervous systems represent, process, and transform information.

Working in robotics, and unaware of the assumptions in interactional linguistics, de Jaegher & Froese (2009) come nevertheless to similar conclusions, namely that there is a relation between interaction and meaning-making. They argue that these two aspects of human agency, namely individual cognition and individual interactions, are interlinked and if our goal is to model human's cognitive capacities, they both must be taken into account.

Jeannerod (2003, p. 1) explains that we recognize ourselves as different from other people by understanding that we are the agent of a behaviour. This sense of agency "is the way by which the self builds as an entity independent from the external world." It follows that self-recognition depends on distinguishing between our bodies that produce actions and actions produced by other agents. This change in the notion of agency is defined by Ibnelkaïd (2019) as "distributed agency", which covers a complexified interactional multimodality, giving rise to a communicative gesture that becomes trans-subjective. Here, the relation to IDN can be seen in the examples involving different types of causality (e.g. mutual, cyclic, spiraling…), but also in the documentation of emergent properties of human interaction.

**Human-Computer Interaction**

Human-computer interaction (HCI) is a multidisciplinary field of study focusing on the design of technology, in particular, the interaction between humans (the users) and computers (Caroll, 2022; Dix, 2022). Interaction is here considered as a multifaceted concept, that covers the interaction (in communication terms verbal, visual, haptic, olfactory) between a human and a machine, between two humans through a machine, one human and an artificial agent through a machine (i.e a sales or information chatbot), or groups of humans and/or agents communicating with each other through a machine (i.e. a massive multiplayer online game, or social media networks). In HCI the terms "interaction" and "interactivity'' are closely related, though there is little agreement over the meaning of the term "interactivity", but most definitions are related to interaction between users and computers and other machines through a user interface.

The Association for Computing Machinery (ACM) defines human–computer interaction as "a discipline that is concerned with the design, evaluation, and implementation of interactive computing systems for human use and with the study of major phenomena surrounding them. …. Because human–computer interaction studies a human and a machine in communication, it draws from supporting knowledge on both the machine and the human side. On the machine side, techniques in computer graphics, operating systems, programming languages, and development environments are relevant. On the human side, communication theory, graphic and industrial design disciplines, linguistics, social sciences, cognitive psychology, social psychology, and human factors such as computer user satisfaction are relevant. And, of course, engineering and design methods are relevant." (Hewett et al; 2022).

As IDNs are about digital artefacts in a technical system, (concerning digital content generation, access and maintenance tools), the interaction concepts of HCI are necessarily to be considered.

**Interactive Digital Narratives**

In the field of interactive digital narratives, Koenitz (2023) distinguishes between two kinds of interactions within an IDN, namely "Interactivity 1" and "Interactivity 2". Interactivity 1, covers the personal interpretation of an artefact and hence refers to cognitive and interpretative acts that take place when engaging with all texts. On the other hand, interactivity 2 is the one typical of digital media where an interactor plans an action and executes it after having considered all the options made available by the system they are interacting with – with the purpose of seeing the system's reaction to it.

These stimuli are (generally, with minor exceptions) ontologically dissimilar: "the IDN represented by the machine actively responds to the physical inputs coming from the player and the player actively responds to the sensory outputs coming from the computer" (bellini, 2022). The engines governing IDN require these mutual actions in order to instantiate the narrative in a specific way among the often countless ones.

Interactions with an IDN are traditionally considered meaningful when they have an impact on the development of the story, when the interactor is afforded "dramatic agency" (Murray, 2017). Dramatic agency is the ability to make 'meaningful choices' and see their effects (Kolhoff & Nack, 2019; Roth & Koenitz, 2019).

# Part II. Interaction in IDN

As IDNs are considered as a system that provides an interplay between programmed discourse strategies and interaction models (System), and means for an interactor to use those (Process) to establish narratives (Product) (Koenitz, 2023), as outlined in Figure 1A..

For the interactors, the process is a double hermeneutic circle in which they reflect both the instantiated narrative path and the possibilities for future interaction (Roth et al, 2018), as outlined in Figure 1B. A similar finding is outlined for the domain of games in Arjoranta, 2022).

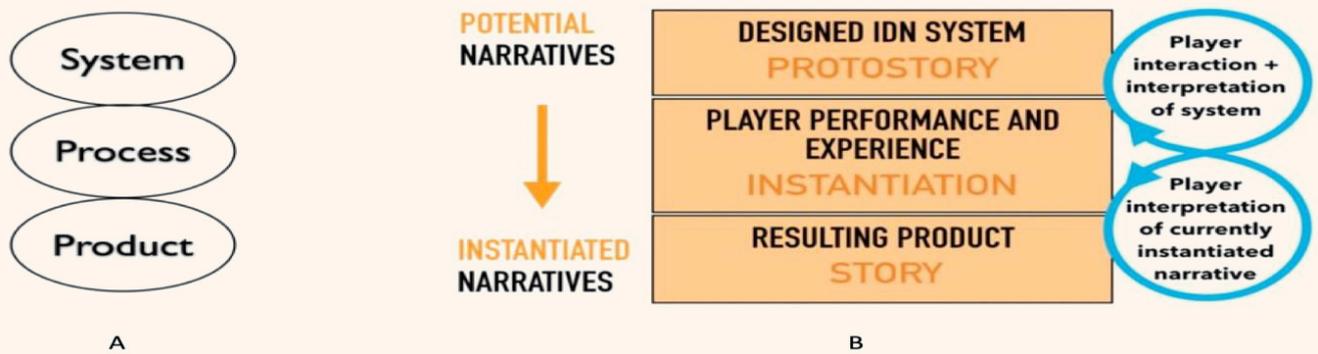

Figure 1: Koenitz' SPP model, where A outlines the relation between the 3 sub-parts of the IDN model, and B demonstrates how the three parts interconnect from the point of view of the interactor/player (adapted from (Koenitz, 2023 for A and (Roth et al, 2018)for B).

Having outlined that the influence of interaction is distributed over various stakeholders within an IDN system, it is now time to look in more detail on the different forms of interaction that can be applied by each role. We start with the creator, then look at the interactors and finally at additional discourse performed on an IDN that can be influenced by various parties.

## The Creator

IDN authoring can be considered as a complicated endeavour that addresses content selection, mode of interaction, audience perception, and narrative generation. Whatever paradigm the authoring process follows, namely a rational (plan-driven) or action-centric (improvised and cyclic) model, there are a set of necessary stages the overall process follows (see for narrative developments (Hardman et al, 2008; Koenitz, 2015; Swartjes &Theune, 2009) and for the development of entertainment software, such as narrative games (de Lope et al., 2017; O'Hagan et al., 2014)). In each of the processes a creator interacts with different sources and potentially other stakeholders in a direct or indirect manner. Inspired by the literature, we consider a 5 process action-thinking authoring model, where each process contributes to the system part in the SPP model:

1. Ideation, where the initial ideas about media production are established.
2. Meaning Making, where a creator specifies the actual message(s) to be conveyed to a particular audience for a particular designed context, resulting in processes where communication strategies in form of articulation techniques are designed and related media assets are captured, generated or transformed.
3. Interaction, where the creator establishes the interaction mechanisms through which the aimed-for audience can access the established narrative space (the protostory) and the related representational and analytic means so that the system can adapt to the specific interactor's needs.
4. Validation, where the creator can simulate the established narrative environment and observe that the aimed-for experience can be achieved.
5. Distribute, where final interaction between end-users and produced media occurs.

The different processes are interconnected and embedded in an overall cyclic and iterative production methodology, where the sequences of analysis, design and implementation are indistinguishable connected. Creators simultaneously refine the mental image of the design object based on the actual perception of the context in and for which it needs to be designed (sensemaking coevolution implementation framework (Ralph, 2010)), which facilitates the critical rethinking of the perceived idea and results in a new design cycle. The aim is to facilitate the creator of an IDN for complex issues to address the interactor's information needs, based on his or her knowledge and skills, so that he or she can actively explore various perspectives, details, and causal relations through the narrative. Thus, the creator is the provider of motivated exploration, reflection and experiences over time, as well as an expectation engineer.

The following outline is oriented towards processes performed by creators of different work fields, such as researchers, journalists, game designers, educators etc, and hence the processes also need to be understood in the context of the respected work flows of the domains those creators operate in. As we cannot address all domains individually, we aim in this whitepaper for generalized role descriptions. We also consider the single creator similar to collaborative work, as the creation steps are rather similar but more synchronization, i.e. interaction between different views, is required when several creators collaborate.

**Ideation**

The interaction here is mainly a self-reflected process that links into current social discourses, personal memories, beliefs, goals, and narrative skills, where notes taking and content collection are predominant approaches to form the idea for the IDN. If the IDN is considered not a work out of interest of the creator but rather as a product for a different stakeholder, then interactions with this organization and potential audience can happen, i.e. in the form of discussions or observations. The ideation process results in an IDN intent (a plan), which in later stages of the authoring process is taken as the basis against which the creator validates the actual development of the protostory space. Considering how the requirement gathering process can make or break IDN development the importance of this development process cannot be overstated and yet, the process itself needs to be adapted to the working needs of the creator (Boehm, 1988; Callele et al., 2005).

**Meaning Making and Interaction**

With respect to the creators' interaction we can distinguish direct and indirect interaction processes. Direct interaction from the point of view of the creator is performed with the authoring tools, the material, the intent, and potential collaborators. Indirectly the creator interacts with the potential audience (see also process 4 - validation), based on the established image, and the stakeholders (i.e. the constituent) through the requirements.

IDN authoring on the level of meaning making needs to be considered as a creative human process that is done in collaboration with a system that establishes potential narratives, where the creator imagines the narrative in multiple iterations, as well as multiple modes, exploring many possible choices and outcomes (protostory). The offered support here is rather service driven, i.e. supporting the finding or the creation of material. This does not mean that a rational, plan-driven model can be excluded, as the engineering of the system requires discrete development sequences, only that the traditional view on those (i.e. pre-production, production, and post production, see also (INDCOR-WP 2, 2023)) and related tasks is not applicable on the whole design process.

As IDNs are digital artefacts, the narrative engineering process necessitates digital tools. Such academic or community tools are widely available, either running on a pc, web, or cloud infrastructure. Good review papers are (Kybartas & Bidarra , 2017; Shibolet, et al., 2018; Kitromili, et al., 2019; EU Cost action INDCOR, 2022), where in particular (Shibolet, et al., 2018) is relevant as it provides a categorisation and description framework for IDN authoring tools (9 categories and 38 descriptors for tool analysis and comparison) based on around 300 tools which have been surveyed and classified (see also (INDCOR-WP 2, 2023) for a detailed discussion of authoring tools).

Tools, as outlined in literature, lean more towards the technical side, as they are basically developed to help non-programmers to build IDN systems. Their design aims at the development of open navigation but their user interfaces and underlying narrative technology in form of templates, modelling, analytics, and rendering mechanisms, are oriented toward the sense-making affordances (syntax and semantics) of the main medium they operate on, most often text or visuals (graphics or video). In complexity, they range between facilitating the integration and organization of recorded content up to fully generative, code-based tools (parsing, graphic rendering, etc), often applied in the domain of narratives in games. The simpler the media and the more reduced the explorative and adaptive capabilities of the envisioned IDN the more creator-friendly the tool. The price to pay is less narrative expressiveness and no or only rudimentary adaptation towards the information needs and skills of the potential interactor.

**Validation**

Validation is an integral part of the design and implementation of IDNs towards content understanding and user agency. (i.e. the user front- and back-end activities) Validating the overall experience of the user is desirable but experience validation is not addressed in most available tools and hence requires intensive interaction of the creator with the system created, in the form of simulations. More details on validation issues can be found in the Whitepaper of WG 3 [2].

## Distribution

With the distribution to the actual user population, either directly or via the contractor, the actual interaction of the creator with the IDN or the development environment stops. Necessarily, the creator can still interact with the IDN, i.e. to fix problems on the system side, but merely he or she will turn into audience or a commenter, thus interacting with the IDN on a discourse level.

## The Audience

The interactive nature of IDNs allows the audience to influence the progression of the story and so their own experience. The main interaction is hence between the audience and the narrative environment. Indirectly, the audience also interacts with the creator, in the form of using and validating the provided means of interaction (on an HCI level) and of adaptation (the system towards the user on needs and skills).

The individual user of an IDN as well as groups of users interact with an IDN in two ways. First, the audience accesses the IDN on a level of getting information and experience-needs satisfied. Here the audience addresses the IDN via front-end processes that reflect the explicit information within a narrative on a moment to moment basis. This can be best described through the ability to make meaningful choices and see their effects which was already discussed in Part 1.

At the same time the audience interacts with the IDN on back-end processes that address the building and maintenance of mental models, which facilitate the prediction on the still available content and the potential means of interaction and system adaptation to idiosyncratic needs. Mainly through the relation between the front- and backend processes the interaction of a user with an IDN creates a process that is shaped by the actions of the user and so results in different instantiated depending on the particular narrative created by the user. The narrative meaning then is a cognitive construct, built by the interpreter in reflected response to the narrative, namely how actions and choices lead to certain consequences and realisation of the importance of agency in an IDN.

The audience reaction to and reception of an IDN product thus depends on factors that might drive enjoyment of the product and other factors that might mitigate its impact and overall experience. Autonomy, presence, flow, character identification and believability, curiosity, suspense, interest, enjoyment, meaningfulness, narrative coherence (making sense and not being confusing) are different elements that the audience can perceive and is rewarded with through interacting with an IDN as discussed in the analysis of a contemporary product, a harbinger in the field, Bandersnatch (Kolhoff & Nack, 2019).

## The Discourse

In essence, the choice of technology and concept can be difficult and it is important that we recognise that there is a real challenge for the domain in guiding the practitioner to best identify the conceptual and technical means and to recognize the relevant working environment. Here INDCOR already provides some initiatives, like the IDN encyclopaedia (INDCOR Encyclopedia, 2024 forthcoming), the collection of IDNs that address complex issues (INDCOR WP-3, 2023), and a descriptive collection of IDN authoring tools[https://omeka-s.indcor.eu/s/idn-authoring-tools/item-set/43].

However, the community has to involve practitioners in a far stronger way as currently done, so that the development of tools are done in a collaborative way between research and the actual working field. As pointed out in the section on "The Creator" the variety of needs is acknowledged but what a good way of addressing this problem is, it's still an issue. Perhaps a sandbox approach might be the direction to go. This means that not one tool for a particular IDN context needs to be generated but rather to aim for an environment where IDN concepts and technologies are mapped onto practitioner profiles that would assist in making the right decisions early on in the development of IDN (Nack, 2023). This requires the collection of adaptable representation templates, argumentation, memory and user models, and IDN analysis tools to be used by a creator. In other words, the IDN domain needs standardisation of not only concepts but also engines, potentially working processes, but far more discussion and exploration is necessary to work in this direction. The community can learn here from the game domain.

Similarly, to the creator's side, a far better understanding of the audience is required. There is, as outlined in Part I, a good understanding of interaction with respect to cognition, embodied interaction, media linguistics (addressing all modalities), and HCI. However, what that actually means with respect to the different levels of direct and indirect interaction and related reflection and experience processes for the interactor is still unclear. Which of the identified representational and related analysis processes are applicable for different IDN aims (i.e. education, exploration, information) applied to different types of either homogeneous or rather divergent audiences? Are effects measurable so that adaptive processes can be implemented? How far would those processes need to be made obvious or hidden to the audience? How do those findings influence the choices made for particular technical modes of interaction?

Interaction with an IDN both mutually enables and constrains cognition. Indeed, interaction is deemed a necessary step to enable cognition in general, and even more so in IDN, in which an interaction is necessary to instantiate the narrative. Cognition is necessary for interaction as IDNs require interactivity 2 of planning and execution, and often foresee successful and unsuccessful interaction strategies from which to choose.

The insights of interactional linguistics can open the way to IDN analytics. The two important assumptions mentioned above can prompt two related observations: if social interactions in the real world are the settings where identities and relationships are shaped, similarly the fictional interactions taking place in the IDN storyworld give shape to the identity of the interactor and their relations with the fictional environment and the characters living therein. On the other hand, forms and practices of linguistic interaction are configured and structured by their context of occurrence, and in a similar way interactors modify their interaction strategy depending on the context that is presented to them in the fictional world of the IDN. Thus, it needs to be further explored how structures of interactional linguistics (text, but also visual media) can be utilised to improve the design, development and perception on representational levels.

Studies on embodiment applied to communication further highlight the complexity of embodied interactions. While embodiment could be deemed as significantly weakened in mediated context such as a digital fictional world, a number of the elements forming real-world complex embodied interactions still hold in IDNs. Among these are languages (with syntax, lexicon, words, sounds, etc.) and other physical semiotic resources (particularly in virtual reality, with gestures, gaze, but also in non-VR artefacts with the manipulation of objects and artefacts), different senses (sight, hearing, touch/haptic sense), contexts, activities and actions, objectives, participants, to whom are attached numerous possible characterisations, such as ongoing social relations, identities, cultures, emotions, etc.

# Conclusion

The white paper showed that interaction is a multifaceted concept that, due to its essential role in IDN, adds with its dynamic contextuality to the overall complexity of IDNs. We outlined briefly some of the relevant sub-fields within the multidisciplinary IDN domain, i.e. interactional linguistics, embodied interaction, cognitive psychology, HCI, and showed in what way they can contribute to the further understanding of the domain in itself, and particular advances that can be achieved for IDN related processes, such as authoring, perception and discourse.

The white paper exemplified what has been achieved with respect to the representation and modelling of the presented concepts and we outlined research directions that INDCOR can still address for the remaining second half of the project duration. It has been made clear, though, that by then mainly the direction of the research path has been established, but that essential work needs to be done after the project will have been finished.